\documentclass{svmult}
\usepackage{geometry}                
\usepackage{graphicx}
\usepackage{amssymb}
\usepackage{epstopdf}
\usepackage{mathptmx}
\usepackage{helvet}
\usepackage{courier}
\usepackage{graphicx}
\usepackage{makeidx}
\usepackage{multicol}
\usepackage{footmisc}
\def\0#1{{\mathrm{#1}}}
\def\1#1{{\mathbb{#1}}}
\def\2#1{{\mathbf{#1}}}  
\def\3#1{{\mathcal {#1}}}
\def\4#1{{\mathsf {#1}}}   
\def\5#1{{{\widetilde{#1}}}}  
\def\6#1{{\overline{#1}}} 
\def\7#1{\breve{#1}}
 \def\8#1{{\widehat{#1}}}  
 \def\9#1{{\check{#1}}}
 \def\fr#1{{\mathfrak{#1}}}

\def\io{{\sc io}}

\def\miss{\mathrel{{\kern3pt\backslash\kern-8.1pt\bigcirc}}}

\def\O+{\bigoplus}

\def\LA+{\;\circ\hskip -9.1pt +\,}
\def\nogo{\mathrel{+\kern-9.0pt + \kern-12.4pt\to}}

\def\v{{\vee}}

\def\xor{{\sqcup}}

\def\Bar{{{}\vrule height 8pt width 1pt depth 3pt{\kern3pt}}}
\def\<{{\left<\right.}}
\def\>{{\left.\right>}}

\def\Grade{\mathop{{\mathrm{Grade}}}\nolimits} 
\def\Grass{\mathop{{\mathrm{Grass}}}\nolimits} 
\def\Rank{\mathop{{\mathrm{Rank}}}\nolimits} 
\def\Dim{\mathop{{\mathrm{Dim}}}\nolimits} 
\def\Diff{\mathop{{\mathrm {Diff}}}\nolimits} 
\def\diff{\mathop{{\mathrm {diff}}}\nolimits}

\def\iso{\mathop{{\mathrm {iso}}}\nolimits} 

\def\poly{\mathop{{\mathrm {poly}}}\nolimits} 

\def\Qi{{\9{\imath}}}

\def\so{\mathop{{\mathrm {so}}}\nolimits}
\def\SO{\mathop{{\mathrm {SO}}}\nolimits} 

\def\spin{\mathop{{\mathrm {spin}}}\nolimits}
\def\su{\mathop{{\mathrm {su}}}\nolimits} 

\def\v{\vee}
\def\ox{\otimes}

\def\BEQ{\begin{equation}}
\def\EEQ{\end{equation}}
\def\BEN{\begin{enumerate}}
\def\EEN{\end{enumerate}}
\def\BI{\begin{itemize}}
\def\EI{\end{itemize}}
\def\BEA{\begin{eqnarray}}
\def\EEA{\end{eqnarray}}
\def\BTA{\begin{table}}
\def\ETA{\end{table}}
\def\BA{\begin{array}}
\def\EA{\end{array}}

\def\fro{\leftarrow}

\def\cto{\;\circ\kern-4pt\to\;}
\def\cfro{\fro \kern-4pt\circ\;}

\def\rar{\rightarrow}

\def\mapsfro{{\fro}\kern-4pt\rule{.5pt}{5pt}}

\def\dar{{\downarrow}}

\def\Diagr#1#2#3#4#5#6#7#8
{\begin{array}{rcccl}%
&              &{\scriptstyle #5}           &                                            &\cr
& #1                                          & \rar& #2                                      &\cr
\cr
{\scriptstyle #8}&\dar \;\; &         &\;\;\dar&{\scriptstyle#6} \cr
\cr
& #4                                   & \,\rar \,& #3                                      &\cr
&                           &{\scriptstyle#7}&  			             &     
            \end{array}}

\title*{Nature as quantum computer}
\author{David Ritz Finkelstein\\
}
\institute{David Ritz Finkelstein \at Georgia Institute of Technology,
Atlanta GA 30342, \email{finkelstein@gatech.edu}}

\begin{document}
\maketitle

\abstract
{Set theory reduces all processes to assembly and disassembly. 
A similar architecture is proposed for 
nature as quantum computer.
It resolves the classical space-time underlying 
Feynman diagrams into a quantum network
of creation and annihilation processes,
reducing  kinematics to quantum statistics,
and regularizing
the Lie algebra of the Einstein diffeomorphism group.
The usually separate and singular Lie algebras
of kinematics, statistics, and conserved currents
 merge into one  regular statistics Lie algebra.}

\section   {Quantum theories}

I once asked Jack Schwartz  what  the difference was between mathematics and physics.
At the time both were just 
equation-juggling  to me.
He was strap-hanging homeward from Stuyvesant High School,
where we had just met, and
he
answered by drawing a hat in the subway air with his free hand:
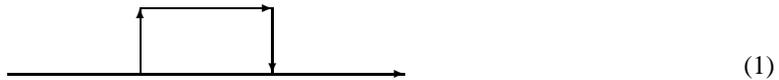
\BEQ
\label{E:HAT}
\begin{picture}(200,25)
\put(25,0){\vector(1,0){150}}
\put(75,0){\vector(0,1){25}}
\put(75,25){\vector(1,0){50}}
\put(125,25){\vector(0,-1){25}}
\end{picture}
\EEQ
He explained that  the bottom line is the real world and
the top line is a mathematical theory.
At its left-hand edge  we
take data from the real world and put them into a mathematical computation, 
and at the right-hand side  we compare the output of the computation with
nature. The loop closes if the theory is right.

This diagram also applies to quantum systems, if
the statistical nature of quantum theory is taken into account.
Then the bottom line is not one experiment on the system but a statistical population of them.

The question remains
of what the symbols of mathematics mean to a mathematician.  
Some decades later I asked  Jack Schwartz
what 	``1" means, and he
replied that it means itself.
This took me aback. 
I had not considered that possibility.
Symbols generally mean something
not themselves.
Memorandum:
\BEQ
1=\mbox{``1"}\/.
\EEQ

After mathematizing aspects
of economics,  robotics,  computing, relativity, 
and even knots,  with a theory that predicted
when ropes would slip on capstans,
Jack Schwartz turned to the problem of putting
quantum theory on firm foundations. 
His concern was not with 
mathematical rigor but physical.
In a paper on ``the pernicious influence of mathematics on science", 
obviously conversing with Wigner's  famous lecture at the Courant Institute on 
``the unreasonable effectiveness of mathematics in the   natural sciences" \cite{WIGNER1960},
Schwartz wrote:
\begin{quote}
The mathematical structure of operators in Hilbert space and unitary
transformations is clear enough, as are certain features of the
interpretation of this mathematics to give physical assertions, particularly
assertions about general scattering experiments. 
But the
larger question here, a systematic elaboration of the world-picture
which quantum theory provides, is still unanswered. 
Philosophical
questions of the deepest significance may well be involved. 
Here
also, the mathematical formalism may be hiding as much as it reveals.
\cite{SCHWARTZ1966}\end{quote}

I respond to this call here, though my answer might not be acceptable to him.
He takes it for granted 
that quantum theory  provides a mathematical world-picture,
faithful or not,
as classical physics did.
It is not clear how literally he intended this.
Some writings on physics
assume that there are  complete world pictures;
in G\"odel's sense of deciding all well-formed questions,
not Bohr's weaker one, of answering all physical questions that can be answered. Bohr's ``complete" is   von Neumann's ``maximal".
 One of the critical differences between quantum and classical physics
 is that quantum physics denies the existence of complete world pictures,
yet  asserts the existence of merely maximal ones.
 Perhaps mathematics 
is the most ``pernicious influence"  when it has been the most ``unreasonably effective."
 
Classically, a mathematical model of a physical system is an isomorphism 
between 
physical predicates about the system
and mathematical predicates about the model.
Physical predicates are defined 
by physical  processes of filtration or categorization.
The stock example is a polarizing filter.

Boole already defined predicates by ``acts of election".
Mental predicates were
mental acts in his theory;
physical predicates are  physical acts in quantum theory.
There are then both input and outtake
predicate logics, corresponding to input and outtake filtrations.
These  are
dual lattices.
They may be operationally defined as
the
Galois lattices of the relation ``Inputs through filter $A$ do not 
trigger counters following filter $B$". 

Mathematical predicates, however, are defined axiomatically
and form Boolean predicate lattices by fiat.
Since models  have Boolean predicate lattices and quantum systems have projective predicate lattices,
the two cannot be isomorphic.
Quantum systems do not have exact mathematical models
because they have different logics than models.

Here the pernicious influence of mathematics is
the highly infectious conviction that there must be an objective public reality,
despite empirical evidence to the contrary.
The mathematics of the quantum theory was invented
 shortly before the quantum theory was born,
demonstrating the unreasonable effectiveness of mathematics.
The quantum evolution has made mathematics more important for physics,
not less.
Mathematics  still provides classical models for classical theories
and  now it also  provides statistical models for quantum systems,
in the following sense.

In classical mechanics a ``state" completely
describes the system,
answering all experimental questions
about it.
Plato said clearly that what is real must have an objective state.

Quantum thought relinquishes this idea. 
This is no great loss, since
we never came close to having the state
of any physical system.
When theorists speak of the state of  Mars, 
they often mean its position and momentum,
or possibly its orientation and angular velocity as well,
ignoring trillions of coordinates of Martian atoms.

Some  give the word ``state" a new statistical meaning
 that can serve quantum theories,
but the result has been continuing widespread confusion. 
We should keep the old meaning of complete information 
for ``state" until the dust settles,
so we can tell our students something important:
 quantum systems have none.

A {\em pure} population is
one that cannot be imitated by mixing 
statistically distinguishable ones.
It goes with an atom of the lattice of predicates.
Classically,  a pure state, when such exist,  is described by a probability distribution 
supported by one point:
 pure implies complete.
A classical  point particle  with continuous coordinates, however,
already has  no pure state,
 because  a point in a homogeneous continuum has  probability 0.
 
In quantum theory,  all systems admit pure probability distributions,
represented by
points in a projective geometry,
of probability measure 1, not 0; and
in  analytic projective geometry, 
by  a ray $\{\lambda \psi\}$ in a linear space $\3V$ 
 that we
associate with the system.

Von Neumann was drawn to projective geometries
that have a continuous range of dimensionalities.
They can have no singlets.
This ignores the problem of the infinities, of which
singlets are the solution.
Classical space-time already has no singlets.

As Malus found for linear polarization, 
the transition probability from one predicate ray to another is 
the squared cosine of the angle between them,
almost never 0 or 1.
Heisenberg therefore called
a vector with this probabilistic interpretation
 a ``probability vector".
Its components are probability amplitudes,
their squares are relative probabilities.
Quantum logic is a square root of classical logic.

But 
a lone classical system has no  probability distribution,
nor has the individual quantum system.
Schr\"odinger and many others
believed, at least at first,
 that a vector $\psi$, up to a factor, 
  exists and evolves in a single quantum experiment, 
 and that it was the state of a real physical object,
 a wave running around the atom.
 For brevity, call holders of this ontological interpretation ``wavers", 
 and holders of Heisenberg's more pragmatic one
 ``chancers".
 Since the state of a real wave is indeed a  wave-function.
the term ``wave-function" is a Trojan horse,
smuggling  waves into the  camp of the chancers.

On the other hand, 
a population of similar 
experiments has at least two  probability vectors or distributions,
 one for the input operation and a dual one for the outtake operation.
 To cope with this ancient duality 
 wavers  say that ``the state vector collapses"
from one to the other
 during a measurement.
This locution is not part of Heisenberg's 
or Kolomogorov's probability theory, for
a probability distribution or vector, be it
classical or quantum,
is unaffected by what happens  to one
 member of its statistical population.
It is not necessarily part of von Neumann's system either,
since it was contributed to his book on quantum mechanics by a friend,
and
does not seem to occur in his later writings.
When we interact with one quantum in a statistical population,
 we may wish to transfer the quantum  to another statistical population
  but
neither probability distribution changes; the quantum does.
   
 Quantum theory eliminated what had been taken for granted: 
 the possibility of a mathematical model of the physical system.
Some classical  conceptions assumes a complete picture of nature, 
as though
taken from 
a preferred viewpoint outside nature.
The corresponding quantum conception would be a plurality of 
partial representations of interactions with small parts of nature. 

It is not obviously possible to visualize an atom completely,
since on the atomic scale photons are not immaterial messengers  
but massive projectiles.
Receiving them changes  us,   so
emitting them must change their sources
as much.
On the atomic scale this reaction by the atom is  greater
than the action upon us is in the human scale.
The smaller the system, the more its perception
exhibits  hysteresis, memory, non-commutativity.
Moreover, we see one atomic transition by absorbing one photon
that we cannot share.
Quantum perception is ultimately private as well 
as non-commutative.

Yet all these impediments to depiction could have  classical models.
They make room for quantum theory, but they do not determine its 
specific features.
If it is obvious that complete descriptions are impossible,
it is  amazing that maximal descriptions are possible.
Partially order classes (predicates)  by proper inclusion, 
and call classes that are $n$ steps away from the empty class,
$n$-plets.
The points of the projective geometry 
are the singlets the maximal descriptions of the quantum theory.

A classical  doublet
includes exactly two singlets.
It cannot have a continuous symmetry group, only $S_2$.

A quantum doublet 
has a theoretical infinity of statistically distinguishable singlets
and an $\SO(2)$ symmetry.

Before studies of polarization, no physicist came close to a singlet of any system;
Newton prepared polarization singlets with his crystals of Iceland spar.
It is likely that when we
reach the bottom of the world,
we will find polarizations and spins on the beach;
not strings,
which have no singlets.

Heisenberg called his probabilistic brand of physics non-objective;
it does not represent objects but 
laboratory actions and 
their probabilities.
In the language of categories,
a classical system can be  presented as a category,
 whose objects are its states.
There is a category of quantum systems too, 
but one quantum system is not a category,
precisely because
 it has not enough objects (states, identities) in the categorical sense.
 Instead it is  represented by an operator algebra, and only statistically.

``Philosophical questions of the deepest
significance" are indeed involved.
Jack Schwartz agreed with Kolmogorov
that probability was not objective,  that  physics ought to be objective,
 and that therefore probability had no role in
a fundamental physical theory.
Some physicists who use quantum theory to great effect
declare that they do not understand it,
and expect it to devolve into a more objective theory.
This likely results from
a deep philosophical preference for objects, 
which are supposed to be knowable ``as they are".
This may be a pernicious influence of the unreasonable success
of classical mathematical models in astronomy.
 But it might be innate,  if
we are hard-wired to see objects.

Exercises in physics often give a complete mathematical
model of the system.
This does not prepare the student for quantum physics.
No one has ever encountered  anything near a complete description 
of any physical system, classical or quantum.
If quantum theory is right, there are none.
Those who study physical systems 
only through such mathematical models
may
find  it absurd and incomprehensible
to say that
physical systems have none.

Again, some note that quantum theory is ``merely" instrumental,
 and find it unsatisfactory on that ground.
This conveys more about the critic's philosophy
than about the quantum theory.
A Beethoven score too is ``merely" instrumental.
Physics is a performing art, 
and physicists are the performers, not 
the spectators; experimenters,  not observers.
The relation of a physical theory to physics is that of a menu to a meal.
It is natural but naive to think that anything in nature has 
a complete objective description;
naive in the sense that throwing a bottle at a villain on a movie screen is naive.

Again, Heisenberg has been criticized by wavers for providing no
mathematical description of the measurement process 
in his theory; despite the fact that
his main point is that none exists.

 We can view any regular quantum system and its co-system
 as a quantum digital computer and its user.
The quantum universe as computer system I sometimes
 call Qunivac for short.
Qunivac differs crucially from artificial computers, however:
It has no fixed hardware;
it is all  quantum jumping.
It includes both computer and
 user. 
 The interface between them  is relatively fixed in artificial computers, but highly movable
 in Qunivac.
  Science is a Promethean attempt to hack into
Qunivac.

\subsection{Canonical quantum theories}

Classical momentum and position coordinates commute: $pq-qp=0$. 
Classical physicists were never aware of this as a physical hypothesis;
it seems to have been an unconscious assumption.
We first became conscious of this assumption when
Heisenberg corrected it to
$\9p\9q-\9q\9p=i\hbar$.
This has led physicists to make other unconscious assumptions
 conscious, and test them.

One operational meaning of this non-commutativity
is that   filters  defining predicates about 
$p$ and $q$
do not commute.  
Such quantum or non-commutative logic was 
found in the laboratory by Newton 
for polarizations of light corpuscles, and
was described in a  quantum-theoretical way by Malus,
who
unwittingly used a two-dimensional real Hilbert space of linear polarizations.
When Boole first axiomatized 
what eventually became Boolean algebra,
he noted excitedly that such a non-commutative logic was possible,
without bringing up Newton's polarizers.

\subsection{Regular quantum theories}

In kinematics we represent
our physical operations on the system
by operators on a space of probability vectors.
A {\em regular} quantum system is one with
a finite-dimensional  probability vector space, 
like a spin or a system of spins
\cite{BOPP1950}. 
Its Lie algebra is simple.
Its observable or normal operators have finite spectra.
A regular theory still mentions infinities, 
such as the real number system $\1R$,
but these result from regarding the co-system as infinite and
are harmless if we abstain from questions about
 the co-system.

\section{Yang space-times}
\label{S:SPACETIMES}

Quantizing space-time to avoid infinities was proposed in about 1930 
 by Heisenberg, Ivanenko, and others.
 The first example was provided by Snyder in 1947 and its commutation relations were immediately
 simplified  by Yang to those of $\mathfrak y:=\so(5; 1)$, 
 precisely
for the sake of simplicity \cite{YANG1947}. 
A {\em Yang space-time} (in the general sense)  is one whose orbital variables span a semisimple Lie algebra, called the Yang Lie algebra.

Yang $\so(5,1)$ is not conformal $\so(5,1)$. 
It
defines a quantum space, while conformal so(5; 1) acts on a classical one. 

Snyder and Yang did not complete their regularizations but continued to represent
their Lie algebras  in the singular $\su(\infty)$ of Hilbert space.
 
Earlier,  R. P. Feynman had quantized space-time
by replacing continuous space-time coordinates by sums of Dirac spin operators (apparently unpublished),
which also leaves Hilbert space;
though he
broke off this work in an early phase to study the Lamb shift for Bethe.
Feynman quantized space-time but not 
momenta. 
The Yang model  quantizes space-time and momentum-energy, but
is still singular.
The Penrose position vector $\2x$  is a finite sum of Pauli spin operators;
the momentum vector is not represented.

Here are the quantum space-time variables of Feynman, Yang, and Penrose, in quantum units:
\BEQ 
 \label{E:QSPACETIMES}
 \begin{array}{lllllll}
\mbox{Feynman  \cite{FEYNMAN1941}}&  \delta \9x^{\mu} &=& \gamma^{\mu}\/,&\cr
\mbox{Yang \cite{YANG1947}}&\9x^{\mu}&=&L^{5{\mu}}= i\eta^{[5}\partial^{{\mu}]},
& {\9p_{\mu}}&=& L_{6{\mu}}=i\eta_{[6}\partial_{{\mu}]}. \cr
\mbox{Penrose  \cite{PENROSE1971}}&  \delta \9x^k&=&\sigma^k, &  
\end{array}
\EEQ

 In such theories, particles do not define 
 irreducible unitary representations of the Poincar\'e
group as Wigner proposed, 
 but irreducible normal representations of a slightly different simple Lie algebra.
 The physical
constants of the Feynman or Yang groups are the speed and action
units  $c,\hbar$ of earlier group de-contractions,  with
 additional elementary time and energy units $\4X$ and
$\4E$ with $\4X \4E = \hbar$, and a huge integer $\4N$. 
The quantum units $\4X$ of time and $\4E$ of energy are here called the chrone  and the erge;
it is not yet clear whether these are the Planck units.
 Under the Yang relativity group, time in chrones in one frame is
just energy in erges in another;
time converts into energy, mass.
The conversion factor is a huge quantum of power, 
one erge per chrone, perhaps the Planck power.
The conversion is not easy but
requires melting the vacuum organization that distinguishes 
energy and time.

The synthesis described here is finite-dimensional, quantizes all the orbital and
field variables, and includes spin.
The other internal coordinates of the Standard Model
are easily tacked on,
but it would be disappointing if the quantization of space-time did not 
lead us to a deeper synthesis of the internal variables.

Einstein's local equivalence principle again suggests that space-time must
be quantized. 
Following Galileo, 
Einstein equates a gravitational field and an acceleration.
 Since the field is quantized, so is the
acceleration.
This is a second time derivative of spacial coordinates with respect to 
time;
if the space and time coordinates were commutative,
the acceleration would be too.
 
Since $i$ and $-i$ are interchanged by Wigner time reversal, they can be
regarded as two values of a discrete dynamical variable $i$ that happens to be central.
This centrality makes the Heisenberg Lie algebra singular and leads to infinities.
The Yang simplification suspended the centrality of $i$.
It is therefore a real quantum theory of the 
St\"uckelberg kind \cite{STUECKELBERG1960}. 
For correspondence with the standard complex theory, Yang provides
 a quantized imaginary $\Qi\in \1S$ 
whose classical correspondent is $i$, with $ \Qi \;\cto i$,
but a self-organization must be invoked to single one
$i$ out of many possibilities.

The Standard Model spinlike groups
can all be defined by their actions on a probability vector space of
about 16 dimensions.
The orbital group seems to act on a much
larger number of dimensions
$\4N\gg 16$.
Therefore events of history are not randomly scattered but highly
organized locally into something like a thin truss dome of four longitudinal dimensions and several short transverse dimensions.
This dome must  support the particle spectrum,
sharp bands of highly coherent transmission, 
and so is presumably crystalline,
as Newton inferred from
transverse photon polarization.

Regular space-times call for regular Lie algebras.
Here is one suggested by those of (\ref{E:QSPACETIMES}),
based on the contraction $\spin(3,3)\cto \fr{hp}(3,3)$:
\BEQ
 \label{E:QSPACETIME}
 \BA{ccrcl}
 \9x^{\mu}&=&L^{5{\mu}}&=& \6{\psi}\,\gamma^{[5\mu]}\,\psi,\\
{\9p_{\mu}}&=& L_{6{\mu}}&=&\6{\psi}\,\gamma_{[6\mu]}\,\psi,\\
 \9 i&=&\4N^{-1}L_{65} &=&  \6\psi\,\gamma^{\top}\,\psi\/.
\EA
\EEQ
Here $\psi$ is a chronon \io\ operator and $\6\psi=\beta \psi$ is 
its Pauli adjoint.
The combination $\6\psi\dots\psi$ is the usual covariant accumulator,
summing many replicas of its argument with attention to polarity.
The eight $\gamma^{\nu}$ generate the Clifford algebra of $\spin(4,4)$, 
with top (volume) element 
$\gamma^{\top}=\gamma^8\dots \gamma^1$.

\section{Whither physics?}

 First the ancient axioms of  space and time failed us in physical experiments
and then
the axioms of Boolean logic.
There are now well-known physical geometries and lower-order physical logic side by side with
the older mathematical ones.
Higher-order logic is evidently next in line
to join
the empire of the empirical.
 
Define an operational theory of a system as  a semigroup 
whose elements are
the feasible operations on the system
by the co-system (the rest of the cosmos, including us),
provided with  their probabilities.
The kinematics gives the possibilities,
the dynamics attaches probabilities.

In quantum theories,
each operation is represented by a projective transformation 
of a specified projective geometry whose 
points are  singlet quantum input or outtake operations.
A special  conic section in the projective geometry
defines transition probability amplitudes.

The first-order logic of the system is the sub-semigroup of filtration operations.
The set theory of the system deals with operations of 
system assembly and disassembly.

Our operations ordinarily 
rely on the organization of the co-system,
natural or artificial,
as do operating manuals or cookbooks.
Highly organized complex elements of the co-system
may enter the system theory only 
through several of their many parameters.

Thus the idea of a Universal Theory is 
absurd. 
Measurements
do not bring us ever closer to Truth, but
invalidate earlier facts as fast as they  validate new ones,
and omit much about the co-system.
A dynamical law 
cannot be universal if it is overridden whenever we measure anything.

Then what  shapes our course?
Two processes by which physical theories evolve 
correspond to
biological evolutionary processes studied by Charles Darwin and Lynn Margulis.
 The Darwinian one is
implicit in  Yang space-time and explicit in work of Segal: 
{\em Physical theory evolves towards semisimple Lie algebras}
 \cite{SEGAL1951}.

Almost all Lie groups have regular Killing forms. 
A singular Killing form is a very rare fish;
the least change in its structure tensor can regularize it.
Singularity
is structurally unstable.
As measurements of structure constants improve, a singular Lie algebra
  has survival
probability 0 relative to its regular neighbors, which outnumber it $\infty$-to-1.
This is the Darwinian argument for semisimplicity.
  
Gerstenhaber, influenced by Segal, described homologically a rich terrain
of Lie groups connected by contractions that carry groups out of stable valleys
of simplicity, along ridges between the valleys, and up to singular peaks
  \cite{GERSTENHABER1964}.
According to the simplicity principle, physics today is  a brook flowing
down the simplicity-gradient to some green valley in Gerstenhaber-land. 

  Thus the Galileo
Lie algebra lies on a singular ridge between the valleys of 
Lorentz  $\so(3, 1)$ and 
Euclidean  $\so(4)$.
Poincar\'e  $\iso(3; 1)$ perches on a higher singular ridge between the
valleys of  deSitter $\so(3; 2)$ and $\so(4; 1)$.

The special relativity Lie algebra
$\iso(3,1)$ makes the observer a rigid body 
with 10
degrees of freedom, like a  speck of diamond dust.
The general relativity group
$\Diff$, however, makes
the observer  an infinite squid, unaffected
by gravity, crossing all horizons freely,
continuously deformable without limit.  
This surely overcompensates.
It still
works at the level of astronomy, and 
it greatly influenced the Weyl,  Yang-Mills,  Schwinger,
and Standard Model
theories of gauge.
Yet general relativity is much more singular than special relativity,
and so  disproves the  Segal theory of evolution as regularization.

The Margulis evolutionary process is
expressed  in biology by  symbiogenesis \cite{MARGULIS2002}
and  in technology by modular architecture \cite{SIMON1962}.
In modular evolution, two modules of some complexity
unite with each other
to form a more complex system that
can use survival strategies of both. 

To cope with complexity, physics has united
 singular modules rather than wait for regular ones.
Stability and complexity are both vital for theories,
and they pull in opposite directions at present.
They must be harnessed to pull together.
General covariance cannot be right in its present form.
Its group $\Diff(4)$ too must be de-contracted.

The de-contraction of $\Diff$ should be compatible 
with the Yang de-contraction of the Heisenberg-Poincar\'e
Lie algebra $\fr{hp}(3,1)\cfro Y$.
Here $Y$ is a high-dimensional orthogonal representation 
$R: \fr y \to \so(N)$, $Y=R(\fr y)$, 
of the simple Yang Lie algebra $\fr y$.
The most
obvious candidate for the de-contracted $\diff$ is some
$\so(N)$, assuming a commutative diagram of Lie algebra 
transformations
\BEQ
{\begin{array}{rcccl}%
&              &           &                                            &\cr
& {Y}                                  & \,\cto \,& {\fr{hp}(3,1)}                                     &\cr
\cr
{}&\dar \;\; &         &\;\;\dar&{} \cr
\cr
& \so(N)                                          & \cto& \diff(4)                                      &\cr

&                           &{}&  			             &     
            \end{array}}
\EEQ

Gauge theories like gravity theory combine identical gauge modules
 at every point of space-time.
 This is a quantification too.
It brings together three Lie algebras: 
 a kinematic one for orbital variables,
a statistical one for the gauge vector quanta, 
and a gauge Lie algebra
for conserved currents.
All need regularization.
In a quantum set theory, all operations are reduced to assembly and disassembly.
Kinematics is all statistics. 
The three singular Lie algebras must then become aspects of one
regular one.

\section{Below Hilbert space}
\label{S:AFTER}

Canonical quantization uses an infinite-dimensional Hilbert space 
$\3H$ of probability vectors. 
This space
 is
still too weak and already too large for quantum theory.

Too large, in that
there are infinitely more orthogonal rays in $\3H$ 
than there can be disjoint pure populations in any physical laboratory.
This makes its unitary Lie algebra singular and leads to divergent sums.

Too weak, in that $\3H$  lacks fundamental concepts of
modular structure and interaction.
True, it can represent general actions;
but under close inspection all actions resolve into interactions.
Hilbert space  has to become smarter
to express these.

It does not seem that a quantum set theory closely modeled on classical
set theory
 will work for quantum physics
as classical set theory works for classical physics
 \cite{FINKELSTEIN2011a}.
The main problem is that  set theory formulates ``laws of thought",
not  of physics.
A set is a collection ``thought of as one".
A proton's position and spin are usually united by  bracing, 
as one unites sets,
but presumably they are not held together by our thinking of them ``as one".
Higher-order logic does not have operational meaning
as the first-order logic does.

Regard Cantor and Peano as proposing 
basic vertices $x\in y$ and $y=\{x\}$,
respectively, for the graph (category) of  all mathematical objects.
Quantum theory must replace these ideational vertices by operational ones 
adequate for the non-category of physical processes (\S\ref{S:GAUGE}).
The result does not resemble set theory enough to warrant the name.

\section{Quantification} 

\label{S:QUANTIFICATION}

Quantification is a logical process that turns a  theory of an individual
into a  theory of a multiplicity of like individuals.
Set theory iterates it.
The logician William Hamilton introduced the term
in 1850.
It has two famous quantum correspondents:
Fermi quantification is regular, and
Bose quantification
is singular but is regularized by Palev (\S\ref{S:PALEV}).

Quantization and quantification traverse the same road in opposite directions.
Quantification assembles individuals into an individual of higher rank.
Quantization resolves an individual into  individuals of lower rank.
Quantization  introduces a quantum constant
when it begins
from a classical limit in which that constant has approached 0. 
In quantification a quantum constant provided by the lower-rank  individual vanishes in a singular limit.
A ``second quantization" is well-known to be a quantification, 
but it is also a ``second quantification"
since it follows 
a quantization, which implies a first quantification.
In this project we express all quantizations, 
including Yang space-time quantization 
(\S\ref{S:SPACETIMES}),
as inverse quantifications, 
to arrive at
the modular architecture
of the quantum universe.
{\em All kinematics is statistics.}

The Standard Model uses  bracing or uniting operation $\{a,b,\dots\}$,
at least tacitly, to assemble quanta from
their various conceptual parts: 
orbital, spin, isospin, color, and so forth.
The lowest-order  predicate algebrahas been quantized.
The higher-order set theory rests on the lower-order;
it must  be quantized or dropped out of physics.  
Here we quantize a small part of it and drop the rest.

Cantor intended to represent the workings
of the infinite Mind of God, 
while physicists seek merely to represent the workings
of finite quantum systems.
For finite algebraic purposes, Peano's one-to-one uniting operation
$y=\iota x= \{ x \}$, sometimes written $\6x$ here, will do
as the key construct of a truncated set theory.
$\iota$  turns what it touches into a monad,
a unit set.
Polyads are  built from monads by disjoint union $a\v b$.
For example $\{a, b\}:= \iota a\v \iota b$ is a dyad;
not to be confused with a doublet.
An $n$-ad is a product of $n$ factors; an $m$-tuplet is a sum of $m$ terms.

Uniting (bracing) occurs in many  important constructs of the Standard Model and 
gravity theory. 
It is used to suspend associativity of the tensor product.
For example it associates spin variables with their proper orbital variables.
Again, a hadron is a triad of quarks, and so the triads must be united
to associate their quarks  properly when a pair of hadrons forms a deuteron.
This  hierarchy of unitings is commonly tacit.
It will be hard to do without it.
But bracing violates  operationality as much as (say)
absolute rest.

Let
\BEQ
\1S=2^{\1S}=\exp_2 \1S
\EEQ
designate the group of {\em perfinite} sets (sets that are  
ancestrally finite, hereditarily finite, finite all the way down).
$\1S$  is an infinite $\1N$-graded group
generated recursively from the empty set 1 by 
\begin{itemize}
\item  the monadic uniting operation  $\iota x$;
and 
\item  the dyadic product operation
$x \xor y=x \;\mbox{\sc xor}\; y$, for  the symmetric union or \sc{xor}. 
\end{itemize}
All  $x\in \1S$ obey the Clifford-like rule
\BEQ
x\xor x =1\/.
\EEQ
 $\1S$ is $\1N$-graded by cardinality (``adicity").
$\1S$ is also  $\1N$-graded by rank, the number of nested $\iota$ operations.
$\1S$ also has a product $x \v y=x\;\mbox{\sc por}\; y$, the Peircian or partial {\sc or},
 determined by $\xor$.
$\mbox{\sc por}$ formalizes Boole's original partial-addition operation 
$\dot +$
and obeys the Grassmann-like rule 
\BEQ
x\v x=0.
\EEQ
This 0 is the OM (for $\Omega$) 
of Jack Schwartz's programming language SETL:
a space-filler indicating the intentional omission of 
any meaningful symbol.
It is the semantic 0.

A plausible probability space for quantum sets 
 is a linearized $\1S$,  
 \BEQ
\9{\1S}=\92\,^{\9{\1S}}=\exp_{\92}\9{1S},
\EEQ
the least linear space that is its own Clifford algebra.
It is
generated recursively from the linear space $\1R\subset \9{\1S}$ representing the empty set
by three operations: 
\begin{itemize}
\item  a monadic uniting operation  $\iota: \9{\1S}\to \9{\1S}$;
\item  a  dyadic Clifford product
$x \xor y$, sometimes written $xy$, for  the  symmetric union; and
\item the dyadic addition operation $x+y$ for quantum superposition
\end{itemize}
 $\9{\1S}$ also has a Grassmann product $x \v y$ determined by
 $\xor$.
 For all $x$ in a certain basis called classical, $x\xor x=\pm 1$ and 
 $x\v x = 0$, 
 as in the classical theory and with the classical meanings.
 
 $\9{\1S}$ is $\1N$-graded by a cardinality operator $\Grade$
$\9{\1S}$ is also graded by the operator $\Rank$, the number of nested $\iota$ operations.

A physical theory needs only a finite-dimensional probability tensor space,
but it is convenient to keep $\9{\1S}$ infinite-dimensional so that
it also contains the singular limits presently in common use.

\section{Palev statistics}
\label{S:PALEV}

Palev regularized quantum statistics  \cite{PALEV1977}  as
 Yang  regularized quantum kinematics \cite{YANG1947}:
by de-contracting a singular Lie algebra into a nearby regular one.

In an even statistics of {\em Palev type $\mathfrak p$} \cite{PALEV1977},
\begin{enumerate}
\item $\fr p$ is a classical (simple) Lie algebra.
\item The  probability vector space  of the individual quantum is $\fr p$\/.
\item The probability algebra of the aggregate is  $\3P=\poly \fr p$, 
an algebra of non-commutative polynomials over $\fr p$
identified modulo the commutation relations of $\fr p$,
defining a representation of $\fr p$\/.
\end{enumerate}
Palev also considers mixed even and odd statistics, where $\fr p$ is a Lie superalgebra.
In the present instance $\fr p=\so(N,N)$ with $N\gg 1$.
Bose statistics 
is merely a useful singular limit of  an even Palev statistics,
ultimately unphysical \cite{FINKELSTEIN2011a}.

There is no physical boundary between statistics and kinematics
(\S\ref{S:QUANTIFICATION}), 
only a historical one.
It is natural to regularize both at once.
Di-fermions obey  a Bose statistics only approximately,
a Palev statistics
exactly.
The Palev Lie algebra can even be the Yang Lie algebra.

A brief review: The three-dimensional Heisenberg Lie algebra $\fr h(1)$, 
with the singular canonical commutation relation
\BEQ
\fr h(1): \quad [q,p]=i\hbar,
\EEQ
underlies both quantum oscillator kinematics and Bose statistics.
$\fr h(1)$  lies on the ridge between
$\spin(2,1)$ and $\spin(3)$.
Quantum relativity needs an indefinite metric,
so choose the indefinite case in this toy example:
\BEQ
\spin(2,1):\quad [q,p]=r, \quad [p,r]=q, \quad [q,r]=p
\EEQ
To contract $\spin(2,1)$ to $\fr h(1)$,
the variable
$r$ must freeze to a central imaginary 
as the dimension $D$ of the representation goes to infinity:
$r\approx Ni$\/, where $N\to \infty$
with $D$.
Call such a process an ``organized singular
limit" and write, for example,
\BEQ
\spin(2,1)\;\cto \mathfrak h(1),\quad\9q\; \cto q, \quad \9p\;\cto p, \quad \9r\; \cto \4N i\;.
\EEQ
The circle in ``$\cto$"
represents the regular algebra, the tip of the arrow
the singular one, and the connecting line the self-organization, if any, 
and the homotopy that connect the regular to the singular.

\section{Neutral metrics}
\label{S:NEUTRAL}
The metric in the Clifford algebra must be specified. 
To represent in the one space $\9{\1S}$ the duality between source and sink
that each experimenter sees,
the  probability form
should be
  neutral, like that of a quantum space in the sense of 
 \cite{SALLER2006a},
 and like the Pauli-Cartan metric $\beta$ of spinor space.
 To fix the sign convention:  source vectors have positive norm,
 sink vectors negative.

Every finite-dimensional Grassmann algebra, and therefore every subspace
$\9{\1S}[r]\subset \9{\1S}$ of finite rank $r$,
has a natural neutral norm, the Berezin $L^{(2)}$ norm
\BEQ
\label{E:NORM}
\|w\|:= \int d\gamma^{\top} w\vee w=\frac d{d\gamma^{\top}}w\vee w =
\beta w w
\EEQ
where $\gamma^{\top}\in \9{\2S}{[r]}$ is a top Grassmann element,
and $\beta w w$ is the polarization of $\|w\|$.
The Berezin norm is identical to the Cartan norm for spinors.

There is a frame-dependent scalar factor in this top element,
and therefore in the norm (\ref{E:NORM}); but 
all physical quantities,
which are of degree 0 in the norm, so this factor drops out.

The previous rank $\9{\2S}{[r-1]}$ is an isotropic space of this norm,
representing linear combinations of sinks and sources with equal probability.

The norm $\beta$ defines the Clifford product $x\sqcup  y$ on $\9{\1S}[r]$
by the Clifford rule
\BEQ
x\sqcup  x = \|x\|
\EEQ
for vectors (of grade  1) and dual vectors (of grade 
$\Dim \9{\1S}[r]-1$).
Here this is the exclusion principle.

In this quantum set theory as in classical set theory, 
any set can be in any set only 0 or 1 times.
Multiple occupancy is forbidden by fiat.
Sets with even statistics have to be pairs of monads.
This set theory does not acknowledge elementary bosons,
which violate Leibniz's principle
that  indistinguishable objects are one.
Indistinguishable laser photons can be $10^10$ and more.
The {\em grade parity} operator $g(x)\doteq 0, 1$ 
of a set $x$ is the parity of the grade 
(= cardinality) of $x$.
It defines the ``statistics" of $x$;
called even (or bosonic)  if $g=0$, and
odd (fermionic) if $g=1$.

The quantum set theory of $\9{\1S}$ immediately conflicts with
 the Standard Model on the conservation of exchange parity
 (statistics).
For any $x,y \in  \9{\1S}$,
$\iota x$ and $\iota y$ are monadic (of grade 1) and
so is their uniting $\{\{x,y\}\}=\iota(\iota x \v \iota y)$.
But in nature so far, as in the Standard Model,
 a composite of two odd quanta 
is always even.
In nature, composition conserves grade parity 
 but not in $\9{\1S}$.
 The question was not considered explicitly in $\1S$,
 which opted to deal in sets alone,
although other modes of aggregation exist in classical thought.
 Sets are all of odd parity in that they obey the exclusion principle
 $x\v x = 0$.
 The functor Grass works well on the probability space of fermions.
 But its iteration $\Grass^2$ violates the conservation of statistics
 (exchange parity $X=0,1$).
 
Another problem with $\9{\1S}$ as a paradigm is that its
$\iota$ is infinitely reducible.
This has classical roots.
The classical $\iota$ is a formal sum of its restrictions 
$\iota^{(m)}$  to sets of cardinality $m$. 
Correspondingly, the quantum $\iota$ reduces
to a sum
\BEQ
\iota=\sum_m \iota^{(m)}
\EEQ
 of its projections on $m$-adics.
 In the classical basis $\{1_s\}\subset \9{\1S}$,
 the operator  $\iota^{(m)}$ has the matrix elements
 \BEQ
 \iota^{\{s_1\dots s_m\}}{}_{s_1\dots s_m}{}
 \EEQ
 in which for any $s_1,\dots, s_m$,  
 $\{s_1\dots s_m\}$ is a single collective index of one higher rank than any of $s_1,\dots, s_m$.
 The repeated unsummed indices break the linear group,
 but the tensor
 \BEQ
 \iota^{\{s_3 s_4\}}{}_{s_1s_2}:= \delta^ {s_3s_4}{}_{s_1s_2}
 \EEQ
 is invariant under the linear group and reduces to $\iota^{(2)}$
 in the classical basis.
Such $\iota^{(m)}$ are the irreducible vertices of this quantum set theory.
If they occur in nature, their occurrences supply their operational definitions.  
If they do not occur in nature, we have to remove them from the physical theory, 
by burying them in its infrastructure if necessary.

There is also a problem with relativity.
Set theory takes the Eternal view and has a distinguished frame,
while a quantum theory admits only the secular perspectives of many limited experimenters.
 $\1S$ and $\9{\1S}$ both have preferred frames.
 
 Here $\9{\1S}$ will be truncated to two ranks,
the points and links required to build a network of interactions.
The rest of the usual hierarchy of ranks is replaced by a hierarchy of clusterings of clusters 
in the network. 
The points have Fermi statistics, the links Palev.

\section   {Quantum events}

Einstein understood an event to be a smallest possible occurrence, and took
the collision of two small hard balls as an approximation to an event. 
Collisions of much smaller things have been studied since then, 
and they have more internal variables than  Einstein's idealized buckshot, such as spins and  Standard Model
charges.
It is remarkable that
the Standard Model  still uses the same mathematical representation of space-time events 
that Einstein did.
Is this unreasonable effectiveness or pernicious influence?

To infer new classical space-time dimensions
from the  internal quantum numbers is archaic today.
Any classical description is a low-resolution many-quantum description.
Quanta are not born out of continua; continua are assembled quanta.
To be sure, the first quanta were explained by quantization.
Similarly, Swift assures us, the first roast pig was discovered when a barn burnt down,
and for some time, a barn was burnt down for every roast pig.
The continua of string theory and of modern followers of Kaluza are our barns.
Eventually it will be recognized, as Dirac did,  that one can 
 have a quantum without quantizing some continuum.

The charges of the elementary particles have  small discrete spectra,
in stark contrast to the quasi-continuous spectra of position coordinates.
This tells us that the charge degrees of freedom 
do {\em not} organize themselves into quasi-continua
as spins do in the models of Feynman and Penrose.
Conceivably  this is the salient difference between the charges and spin.

According to present physical theory,
we never perceive space-time itself, 
only  quanta.
Quantizing space-time  is best understood as extending the familiar
quantization of orbital angular momentum
to the other orbital variables $x^{\mu}$, $p_{\mu}$,
like Feynman, Snyder, Penrose, and Yang.
It resolves a fine-structure of quanta
that canonical quantum theories smear out.
Since our measurements always concern quanta, 
we have no need for both quanta and  space-time.
Regard space-time as another classical reification,
a mental extension of the solid  laboratory floor beam
to the distant stars.

Particle collisions take the place of Einstein's buckshot collisions.
In the relativistic canonical quantum theory the 15 orbital operators of en event,
\BEQ
x^{\mu}, p_{\mu}, L_{\mu'\mu},i\in \mathfrak {hp}(3,1).
\EEQ
span 
the singular Heisenberg-Poincare Lie algebra.
In addition quanta have the internal variables of the Standard Model.
To describe one quantum in a canonical theory it suffices to tensor-multiply certain of these spaces
and unite the product with braces or $\iota$.
In the canonical theory, however, collisions  break up into several \io\ processes for quanta.
One space-time quantization resolves these in turn into chronons two ranks of aggregation lower,
the fundamental actions
of the theory.

Since fermions exist, 
 we cannot  begin the inductive construction of the quantum universe as computer
  with the empty set alone, 
 which is even.
Begin with  a foundation of $\4N$ primal odd chronons represented by 
basis spinors $\chi_a\in \3X$, $a\in \4N\gg 1$.
A set-theoretic  fermion field hierarchically unites an odd number of chronons first into
events $\epsilon=\{\chi_e \v\dots\v \chi_e\}$, and then into
fields $\phi=\epsilon_a\v \dots \v\epsilon_b$.
To form a gauge boson, unite chronons into di-chronons, into 
gaugeon events, into a gaugeon field.

Set theory envisages the construction of its universe from the empty set in an infinity of ranks,
by a Mathematician outside the theory.
A similar construction within superset theory $\9{\1S}$  requires
on the order of 10 ranks or less to accommodate physics.
It does not represent the entire universe,
since mundane experimenters, like the extramundane Mathematician, 
are there from start to finish, 
controlling the system but largely undescribed in the theory.

Nevertheless it is advantageous for $\9{\1S}$ to be infinite-dimensional.
so that the singular limit of classical space-time
can be carried out within
$\9{\1S}$.

Absolute Space and Absolute Time have left the theater but 
Absolute Space-Time remains in the Standard Model and general relativity.
How did this myth begin?

Etymologically, 
a  ``line" is a linen string, 
a ``point" is a
puncture or stick, 
and ``geometry" is  earth-measurement. 
Sticks connected by
linen strings used to puncture the Nile 
flood-plain each spring.
This is supposed to be the origin of geometry.
These sticks, however,  have position,
momentum, angular momentum, time, and energy. 
Presumably these were first neglected and then lost
on the way to Greece, giving rise ultimately to the constructs
of  space-time.
In quantum field theory  space-time is merely an index on some variables,
part of the infrastructure.
Space-time coordinates are actually carried by physical quanta, 
not by mythical space-time points.

The fundamental events in nature are then  \io\ operations for quanta.
In the Standard Model
every fermion carries many variables:
one hypercharge $y$, 
one generation index $g$, 
three isospin $\tau^k$, 
four Dirac $\gamma^{\mu}$, 
eight color charges $\chi^c$, 
four space-time $x^{\mu}$, 
four energy-momentum $p_{\mu}$, and
six angular momenta $L_{\mu'\mu}$.
Like atomic number and atomic weight, 
these variables tell us something about the structure and composition of the fermion.
We are to  fit them all
into a semisimple operator algebra  $\3E$ of event variables.

To recover the canonical quantum theory from a Yang $\spin(3,3)$  theory,
one freezes one rotational degree of freedom 
$L_{65}\cto Ni$,
as a step in the organized singular
limit in which $\hbar, \4X\to 0$ and $N\to \infty$.
Like the Higgs and gravitational fields, $i$ is the non-zero vacuum value
of a non-commutative field operator $\Qi=\L_{65}/N$ .
Presumably all vacuum values result from freezing and self-organization.
The possibility that the Yang $\Qi$ is the Higgs field has not been excluded.

In the Standard Model, odd probability vectors form a Clifford algebra, 
even ones a Heisenberg Lie algebra,
and orbital  coordinates  $x,p$ another 
Heisenberg algebra of lower rank.
This singular structure is presumably a
jury-rig. 
A nearby finite-dimensional  algebra
arises from the construction out of fermion pairs;
not a Bose Lie algebra but a Palev one
based on $\so(N,N)$.

Since the noncompact Lorentz group will not fit into a compact unitary group,
a finite-dimensional relativistic quantum theory has to renounce
definiteness of the probability norm as well as the space-time norm.
 Relativistic spinor theory provides one resolution \cite{DIRAC1974}.
The space $\4R$ of real Majorana Dirac spinors has no invariant definite
metric form.  
Instead  it has one Pauli form $\beta$ that is invariant but  not definite, 
and a plethora of Hilbert forms  $\4h$ that are definite but not invariant, associated with different time axes;
and nevertheless it works.

Take spinor spaces as elementary building blocks,
so that their aggregates inherit this multi-metric structure.
Correspondence with present physics requires that a unique
invariant global hermitian form $\4H\cfro \4h$ exist
 in the singular limit of classical space-time.

\section{Quantum gauges}
\label{S:GAUGE}

Experience severely breaks the kinematic
symmetry between position and momentum in quantum mechanics.
So does the locality principle,
which requires field variables coupled in the Lagrangian or action to  share a
 space-time point,
 not a point in the Fourier transform momentum-energy 
space. 
So does gauge field theory, including general relativity.
We must resolve this discord between  the diffeomorphism 
and canonical groups 
into a harmony;  they work too well 
 to be merely discarded.

Gauge theories today rest on the following relation,
with readings depending on context;
``energy" stands for ``energy-momentum" here:
\BEQ
\label{E:GAUGE}
\BA{ccccc}
D_{\mu}&=&\partial_{\mu}& - &\Delta_{\mu},\\
 \mbox{covariant derivative} &=& \mbox{Lie derivative} &-& \mbox{connection, }\\
 \mbox{kinetic energy} &=& \mbox{total energy} &-& \mbox{potential energy}.
\EA
\EEQ
In a Yang-Feynman quantum event space,
the invariant concept is the Yang-Feynman  quantized coordinate $\9p_{y'y}$,
This is a cumulation 
\BEQ
\9p_{y'y} =\6{\psi}\,\gamma_{y'y}\,\psi 
\EEQ 
of many spin-matrix terms  $\gamma_{y'y}$, 
generating a regular Lie algebra $\fr y$.
In turn, $\9p_{y'y} $ has a singular semi-classical contraction $p_{y'y}\in \fr {hp}(3,1)\cfro \fr y$, 
the usual 15 canonical quantum event coordinates $p_{y'y}= (x^{\mu}, p_{\mu}, L_{\mu'\mu}, i)$.   
Then underlying the usual gauge field kinematics (\ref{E:GAUGE}) is 
the single-event relation
\BEQ
\BA{ccccc}
\9p_{y'y} &=& p_{y'y} &- &P_{y'y},\\
\mbox{Yang coordinate} &=& \mbox{canonical coordinate} &-& \mbox{quantum correction.}
\EA
\EEQ
Both terms on the right-hand side are singular; only the left-hand side is physically meaningful.
Accumulate this relation over all events in a field by a higher-rank cumulation $\9{\phi}\dots \phi$
and we arrive at the regular correspondent of (\ref{E:GAUGE}).

In older terms:
classical gravitational curvature and the electromagnetic field are
higher order corrections that remain 
when we contract quantum event space
to a flat, field-free, classical space-time.
They are classical effective descriptions of 
quantum non-commutativity at the chronon level.

C. S. Peirce noted that while only line graphs can be
built up from a 2-vertex alone, the most general graph can be simulated with 
a triadic vertex alone.
The natural physical candidate for a universal vertex
 is  indeed triadic,
that of gauge physics, with three limbs:
a chiral spinor $\psi$, its dual $\6{\psi}$,
and a gauge vector boson (actually, palevon)  $\phi$:
\BEQ
\label{E:TRIADIC}
\6{\psi}\gamma^mA_m\psi =: \6{\psi}\phi\psi=\gamma^{y"y'y}\6{\psi}_{y"}\phi_{y'}\psi_y\/.
\EEQ
Here $\spin(8)$ triality cries out for physical interpretation, so far in vain.

The proposed probability space 
consists of all tensors constructed inductively from a finite number of triadic vertices 
$\gamma$ of (\ref{E:TRIADIC}),
by 
tensor multiplication,
linear combination, 
contraction (connecting two compatible lines),
and identification modulo commutation relations
of the Yang-Palev kind. 
Odd lines obey Fermi statistics.
Even lines obey the unique Palev statistics induced by this Fermi statistics.

The outstretched arms of $\gamma$ represent chiral spinors, 
and the  leg
represents a vector in the first grade of a Clifford algebra
associated with the spinor space.
The spinors have exchange parity 1 and the vector exchange parity 0.
All the terminals of $\gamma$ are polarized. 
Spinors plug only into dual spinors, and dyads only  into dyads.

Call the quantum structure whose history probability tensors 
are so constructed a quantum {\em interaction network}. 
Its probability tensors derive from
 Feynman diagrams and Penrose spin networks 
more than Cantor sets.

This diagram algebra will be developed further.

But it begins to seem likely  that the three simple physical Lie algebras introduced by  Yang into  kinematics,
Palev into statistics,  and Yang-Mills into differential geometry  are actually different representations of one.
Whether this shows the pernicious influence or the unreasonable effectiveness of mathematics remains to be seen.

\section{Acknowledgments and references}
Jack Schwartz and Martin Davis 
taught me some of  the logical works of Post and G\"odel.
Roger Penrose and Feynman showed me  
their spin atomizations of space or space-time.
The Lindisfarne lectures of Lynn Margulis
introduced me to symbiogenesis.
Tchavdar Palev discussed his statistics with me.
David Bohm convinced me 
that the mathematics in mathematical physics
is best regarded as an extension of natural language.
Frank (Tony) Smith shared his knowledge of 
Lie and Clifford algebras with me. 
James Baugh,
David Edwards,
 Shlomit Ritz Finkelstein,
 Andry Galiautdinov, 
 Dennis Marks, 
 Zbigniew Oziewicz,  
 Heinrich Saller, 
 Sarang Shah, and
 Mohsen Shiri-Garakani
 provided many helpful discussions and corrections.
 I thank them all deeply.
 
 This paper is dedicated to Jack Schwartz.

\end{document}